\PassOptionsToPackage{dvipsnames}{xcolor}
\documentclass[sigconf,screen,authorversion]{acmart}

\usepackage{xcolor}
\usepackage{hyperref}
\usepackage{acronym}
\usepackage{mathtools}
\usepackage{bm}
\usepackage{multirow}
\usepackage{csquotes}

\usepackage[capitalise,noabbrev]{cleveref}
\crefformat{section}{\S#2#1#3} %
\crefformat{subsection}{\S#2#1#3}
\crefformat{subsubsection}{\S#2#1#3}
\crefrangeformat{section}{\S#3#1#4 to~\S#5#2#6}
\crefrangeformat{subsection}{\S#3#1#4 to~\S#5#2#6}
\crefrangeformat{subsubsection}{\S#3#1#4 to~\S#5#2#6}

\newcommand\free[0]{\vspace{1em}\noindent}

\begin{document}

\author{Nils Japke}
\affiliation{%
    \institution{TU Berlin}
    \department{Scalable Software Systems Research Group}
    \city{Berlin}
    \country{Germany}}
\email{nj@3s.tu-berlin.de}
\orcid{0000-0002-2412-4513}

\author{Furat Hamdan}
\affiliation{%
    \institution{TU Berlin}
    \department{Scalable Software Systems Research Group}
    \city{Berlin}
    \country{Germany}}
\email{fuha@3s.tu-berlin.de}
\orcid{0009-0005-6452-0990}

\author{Diana Baumann}
\affiliation{%
    \institution{TU Berlin}
    \department{Scalable Software Systems Research Group}
    \city{Berlin}
    \country{Germany}}
\email{diba@3s.tu-berlin.de}
\orcid{0009-0006-3000-6691}

\author{David Bermbach}
\affiliation{%
    \institution{TU Berlin}
    \department{Scalable Software Systems Research Group}
    \city{Berlin}
    \country{Germany}}
\email{db@3s.tu-berlin.de}
\orcid{0000-0002-7524-3256}

\title{Investigating the Impact of Isolation on Synchronized Benchmarks}



\copyrightyear{2025}
\acmYear{2025}
\setcopyright{cc}
\setcctype{by-nc}
\acmConference[UCC '25]{2025 IEEE/ACM 18th International Conference on Utility and Cloud Computing}{December 1--4, 2025}{Nantes, France}
\acmBooktitle{2025 IEEE/ACM 18th International Conference on Utility and Cloud Computing (UCC '25), December 1--4, 2025, Nantes, France}
\acmDOI{10.1145/3773274.3774703}
\acmISBN{979-8-4007-2285-1/2025/12}

\begin{abstract}
    Benchmarking in cloud environments suffers from performance variability from multi-tenant resource contention.
    Duet benchmarking mitigates this by running two workload versions concurrently on the same VM, exposing them to identical external interference. However, intra-VM contention between synchronized workloads necessitates additional isolation mechanisms.

    This work evaluates three such strategies: \emph{cgroups} and \emph{CPU pinning}, \emph{Docker containers}, and \emph{Firecracker MicroVMs}.
    We compare all strategies with an unisolated baseline experiment, by running benchmarks with a duet setup alongside a noise generator.
    This noise generator \enquote{steals} compute resources to degrade performance measurements.

    All experiments showed different latency distributions while under the effects of noise generation, but results show that process isolation generally lowered false positives, except for our experiments with Docker containers.
    Even though Docker containers rely internally on \emph{cgroups} and \emph{CPU pinning}, they were more susceptible to performance degradation due to noise influence.
    Therefore, we recommend to use process isolation for synchronized workloads, with the exception of Docker containers.
\end{abstract}

\maketitle

\section{Introduction}
\label{sec:introduction}

Many applications and services nowadays rely on the cloud, which makes the performance of cloud applications particularly interesting to software developers.
However, cloud computing platforms suffer from unpredictable performance~\cite{Performance_Variation_and_Predictabilit}.
Many techniques have been developed to overcome this problem, such as \emph{Randomized Multiple Interleaved Trials} (RMIT)~\cite{AbediOld,abedi2017conducting} and \emph{duet benchmarking}~\cite{inital,Bulej_2020}.

RMIT overcomes the unstable cloud performance by increasing the number of iterations across levels (such as replicating the benchmark using different instances), and shuffling them.
This, however, strongly increases the overall experiment time and cost.
Duet benchmarking has become popular in research as it aims to overcome this problem in regression testing scenarios.
For a regression test, two versions of the \emph{system under test} (SUT) are compared against each other on the same \emph{virtual machine} (VM) in the cloud.
However, because the two SUT versions might influence each other depending on, e.g., CPU scheduling, strong isolation of the processes is necessary.

This leads to the following questions:
\begin{itemize}
    \item How strongly are measurements affected by other processes running on the same VM?
    \item Which different process isolation techniques can help to ensure high measurement quality?
\end{itemize}
The necessity of running the SUT in a controlled environment during a benchmark is generally understood and recommended in order to remove any outside influence on the measurements that we can directly account for.
Prior studies have already focused on quantifying the performance loss to \enquote{noisy neighbors}, which often happens in cloud environments~\cite{volpert2023isolation}.
For a synchronized benchmarking experiment such as duet benchmarking, however, the exact influence on measurement results by outside influence has not been established yet.

In order to answer the research questions posed above, we provide the following contributions:
\begin{itemize}
    \item We develop a noise generating application that attempts to influence performance measurements of a benchmark by using a high amount of CPU time.
    \item We run benchmarking experiments using duet benchmarking with a testbed application first presented by Japke et al.~\cite{Japke_2023} to compare the results with different process isolation techniques at different noise levels.
    \item In particular, we compare the following isolation setups:
    \begin{enumerate}
        \item No isolation
        \item Resource assignment using \emph{cgroups} and \emph{CPU pinning} of the main SUT process
        \item Isolating the SUT in a Docker container (including \emph{CPU pinning} of the container on the host machine)
        \item Isolating the SUT in a Firecracker MicroVM (including \emph{CPU pinning} of the MicroVM on the host machine)
    \end{enumerate}
\end{itemize}

We find that noise levels are strongly apparent when using either no isolation or Docker containers, but are not as clearly visible when using \emph{cgroups} and \emph{CPU pinning} or MicroVMs.
Additionally, we find that when calculating the relative differences of median performance between the SUT versions and using state-of-the-art regression detection techniques, all setups mostly correctly report \enquote{no change}, as expected in an A/A test, except when using Docker containers, which had a higher amount of false positives, even without any noise present.
Therefore, we conclude that the duet benchmarking technique can work even at higher noise levels to provide largely correct results.
However, stronger isolation still helps to reduce the amount of false positives.

\section{Background}
\label{sec:background}

In this section, we introduce the necessary scientific background that this paper relies on.

\emph{Regression Testing.}
Regression testing refers to comparing two different versions of a SUT.
This is typically done by running the same benchmark workload separately against both versions while collecting performance measurements (e.g., latency).
Then, both results are compared using statistical methods, such as comparing confidence intervals or using hypothesis tests.

\emph{Cloud Performance Variability.}
The performance of VMs on cloud platforms typically varies, both over time and even across different VMs of the same type~\cite{Performance_Variation_and_Predictabilit}.
This creates challenges when trying to run benchmarks of applications in cloud environments as different performance characteristics of the platform can bias measurements.

Several techniques have been proposed to deal with these challenges.
Here, we focus on the two following techniques.

\emph{Duet Benchmarking} aims to deal with cloud performance variability without needing additional infrastructure~\cite{inital,Bulej_2020}.
The central idea of duet benchmarking is that for benchmarking experiments with comparative results, such as when performing a regression test between two versions of an SUT, we can perform both benchmarking experiments simultaneously on the same VM.
This ensures that any performance variability affects both SUTs in the exact same way, so that we can still measure the relative performance difference between both SUTs.
Using duet benchmarking, we cannot obtain any \emph{absolute} performance metrics that are usable outside of such a comparison, however, as the single VM might bias the results due to cloud performance variability.
Duet benchmarking also contains an additional implementation concern, i.e., isolating both SUT processes from one another to ensure that both are treated fairly and have access to the same amount of resources during the entire experiment.

Duet benchmarking has been used in scientific studies on application benchmarks and microbenchmarks~\cite{grambow2022using,Japke_2023,inital,Bulej_2020}.
The main advantage over RMIT is the significantly lower infrastructure cost and experiment runtime, but as explained above duet benchmarking is limited to comparative benchmarks such as regression tests.

\emph{Process Isolation.}
We look at the following strategies for process isolation under a Linux host system:
\begin{itemize}
    \item \emph{CPU pinning and cgroups:} With the \texttt{taskset} utility, we can set a process to only execute on one CPU core.
    We combine this with \emph{cgroups}, which are a mechanism of the Linux kernel to assign different resource restrictions to processes, including CPU, memory, and I/O.
    \item \emph{Docker Containerization:} Containers are a lightweight method for isolating processes and libraries in their own separate environment.
    They build on \emph{cgroups} and other mechanisms of the Linux kernel to achieve this, and still run using the host kernel.
    \item \emph{Firecracker MicroVMs:} Firecracker is a Virtual Machine Manager for MicroVMs, which are a lightweight variant of VMs.
    MicroVMs use their own \emph{guest kernel}, and are, therefore, a complete Linux subsystem running on the host machine.
\end{itemize}

\section{Study Design}
\label{sec:study_design}

In this section, we describe our study design.

\subsection{Benchmarking Testbed Application}
The application used as the SUT in this study is a lightweight service called \texttt{flight-booking-service}.
Originally developed by Japke et al.~\cite{Japke_2023} for investigating the capabilities of application-level and microbenchmarks in detecting performance regressions, this service has been adapted to demonstrate the effect of isolation strategies in a duet benchmarking environment.

The application is written in Go and replicates core functionalities of a typical flight booking system.
It includes endpoints for searching flights, retrieving available seats, and booking flights.
It exposes the following HTTP API: \textbf{GET} for \texttt{/destinations},\break \texttt{/flights}, and \texttt{/flights/{flight\_id}/seats}, as well as \textbf{POST} for \texttt{/bookings}.

As part of the testbed application, Japke et al.\ also provide an application benchmark, and a framework to run this benchmark for a regression test using \emph{duet benchmarking}~\cite{Japke_2023}.
The benchmark includes two different workload scenarios:

\free
\textbf{S1 -- Flight Search:} This scenario mimics a user searching for available flights.
It begins with a request for location airports (\texttt{GET /destinations}).
A random location is selected, followed by a query for departing flights from there (\texttt{GET /flights?from=\$airport}).

\free
\textbf{S2 -- Flight Search and Booking:} This extended scenario adds a booking step.
After selecting a flight, the user retrieves seat availability (\texttt{GET /flights/\$id/seats}) and proceeds to reserve seats by submitting a booking request (\texttt{POST /bookings}).

\free
To simulate realistic usage patterns, the two scenarios are executed with different weights.
The k6 configuration defines 50 virtual users (VUs) executing 2,000 iterations each for S1 and 10 VUs executing 380 iterations each for S2.
This results in approximately 103,800 search requests and 3,800 booking requests per benchmark run.

Each benchmark runs for roughly 30 minutes.
Since the focus lies on performance characteristics rather than correctness, occasional timeouts or HTTP errors are tolerated as long as they do not systematically affect measurement outcomes.

\subsection{Noise Generator}
To simulate external interference and evaluate noise resilience, we use a custom noise generator written in Java.
It supports the generation of CPU noise by running multiple threads that execute compute-bound loops.
Users can define the number of threads, maximum CPU usage, and runtime duration.
If a CPU usage cap is set, threads periodically sleep to throttle consumption.

\subsection{Performance Change Detection}
\label{sec:perf-change-detection}
For all experiments, we run duet benchmarking with different process isolation strategies in a regression test using the same version of the SUT twice (i.e., an A/A test).
This follows the established \emph{duet benchmarking} methodology~\cite{Japke_2023, inital, Bulej_2020, rese2024increasing, grambow2022using}.
In order to analyze the results of each experiment regarding measured performance regressions, we use two different statistical methods.

For the first method, \emph{confidence interval overlap}, we first calculate the \emph{relative change} by dividing the median latency of version 2 by the median latency of version 1.
Then, we construct a confidence interval (CI) around the measured relative change with bootstrapping, a non-parametric method to calculate confidence intervals using resampling (confidence level 99\%, 10,000 bootstrap iterations).
Finally, we compare whether the CI overlaps 1, which indicates no change.
If it does not overlap 1, we report a significant performance difference.

For the second method, we compare the latency values of both versions using the Wilcoxon rank-sum test.
This is a non-parametric test, which compares the median of two different samples.
We use a confidence level of 99\%, meaning if the p value is lower than 0.01, we report a statistically significant performance difference.

In order to quantify the influence of noise, we apply both techniques to result data from the noise phase and non-noise phase separately, as well as all data overall from one experiment.
The separation helps us identify the impact of noise influence, while practitioners would only observe the overall data.

\section{Evaluation}
\label{sec:evaluation}

In this section, we present the experimental setup and results of evaluating isolation strategies in duet benchmarking within cloud environments.
The goal is to assess how effectively these strategies reduce performance interference.

\subsection{Experiment Setup}
\label{sec:experiment}

\begin{figure}[ht]
    \centering
    \includegraphics[width=\columnwidth]{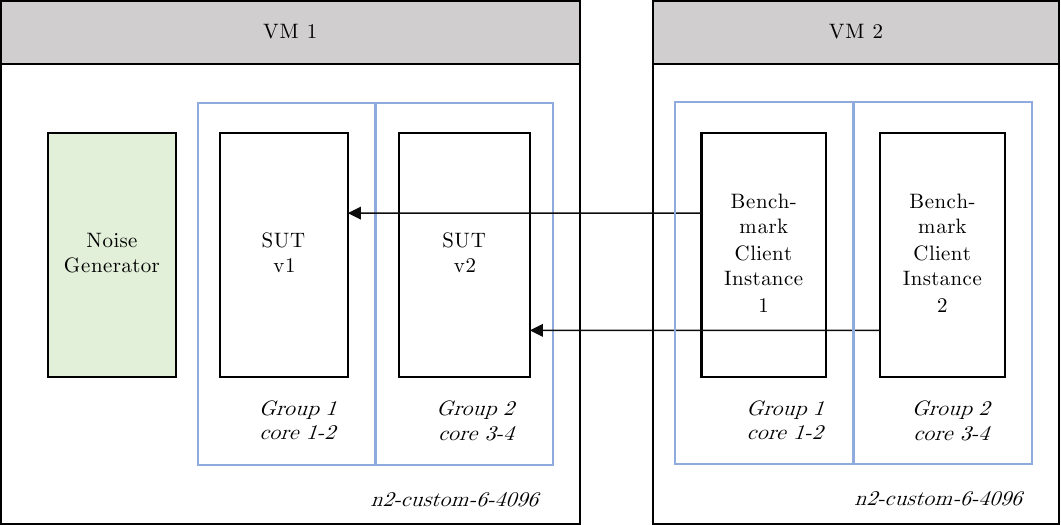}
    \caption{An illustration of the general experiment setup.
    The different \emph{groups} on both VMs represent the different isolation strategies, i.e., using \emph{cgroups} and \emph{CPU pinning}, using \emph{Docker} with similar functionality, and using \emph{Firecracker MicroVMs} with similar functionality.
    In the baseline experiment, we use no isolation strategy, which means that this blue box does not exist for that particular setup.}
    \label{fig:experiment-setup-general}
\end{figure}

We host all experiments using Google Cloud.
Each VM uses the \texttt{n2-custom-6-4096} configuration, with 6 virtual CPUs and 4 GB of memory.
\cref{fig:experiment-setup-general} shows the general experiment setup.
One VM hosts the SUT instances, while a second VM runs the benchmarking clients.
This separation prevents competition for resources between the SUT instances and the workload generator.
Both the SUT versions and the two instances of the benchmarking client are isolated using the same strategy that is used for the particular experiment (potentially unisolated in the baseline experiment).
The noise generator is deliberately left unisolated, as its purpose is to introduce system-level interference that ideally affects both SUT instances equally.
We run four experiments: (1) baseline -- no isolation, (2) \emph{cgroups} and \emph{CPU pinning isolation}, (3) \emph{Docker container} isolation, and (4) \emph{Firecracker MicroVM} isolation.
Each experiment includes six configurations: \emph{0, 3, 6, 20, 40, and 60 noise CPU threads}.
We run each experiment for approx.\ 30 minutes.
The noise generator always starts deterministically \emph{200 seconds} after experiment start, and shuts down \emph{500 seconds} after experiment start.
We also remove the first and last 60 seconds of measurements as warmup and cooldown periods.

We use A/A testing, i.e., we compare the same version of the SUT to itself in a regression test.
This helps us to identify the influence of the noise generator, as we know the ground truth of the regression test is 0\% performance deviation.
This means that any observed performance differences are caused by environmental variability or insufficient isolation.

We provide all code and data artifacts, as well as an online appendix in our replication package.\footnote{\url{https://github.com/njapke/isolation-replication-package}}
Due to the large amount of result data, we only show highly aggregated plots in this section.
For full result tables and supplemental plots, see our online appendix.

\subsection{Results \& Findings}

\begin{figure}[ht]
    \centering
    \includegraphics[width=\columnwidth]{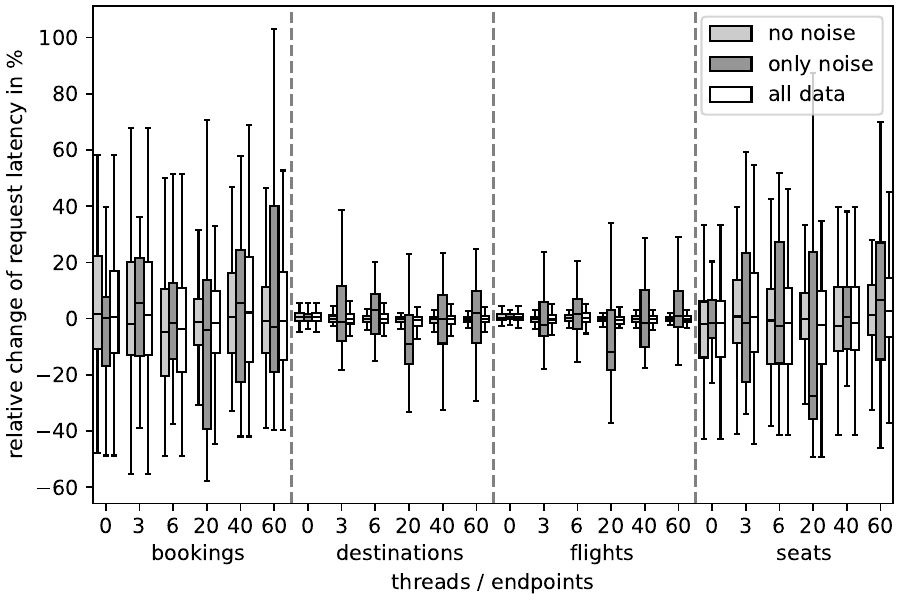} 
    \caption{Boxplots for all results of \emph{Experiment 1 -- Baseline}.
    For each endpoint, we show the distribution of relative changes in request latency throughout the experiment for all configurations.
    We also show the distribution outside the noise generation (\emph{no noise}), during noise generation (\emph{only noise}), and all result data together (\emph{all data}).}
    \label{fig:experiment-baseline}
\end{figure}

\begin{figure}[ht]
    \centering
    \includegraphics[width=\columnwidth]{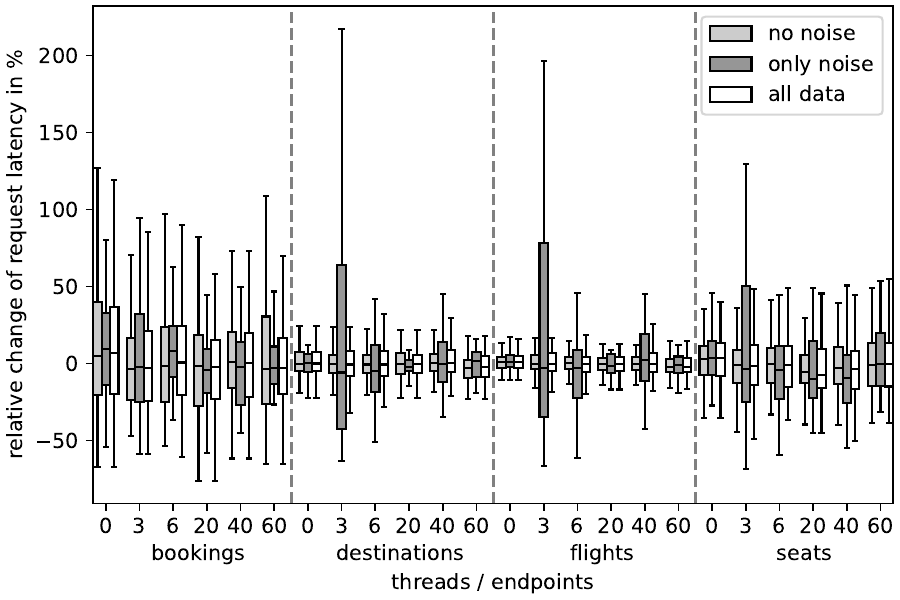} 
    \caption{Boxplots for all results of \emph{Experiment 2 -- cgroups and CPU pinning}.
    For each endpoint, we show the distribution of relative changes in request latency throughout the experiment for all configurations.
    We also show the distribution outside the noise generation (\emph{no noise}), during noise generation (\emph{only noise}), and all result data together (\emph{all data}).}
    \label{fig:experiment-core-isolation}
\end{figure}

\begin{figure}[ht]
    \centering
    \includegraphics[width=\columnwidth]{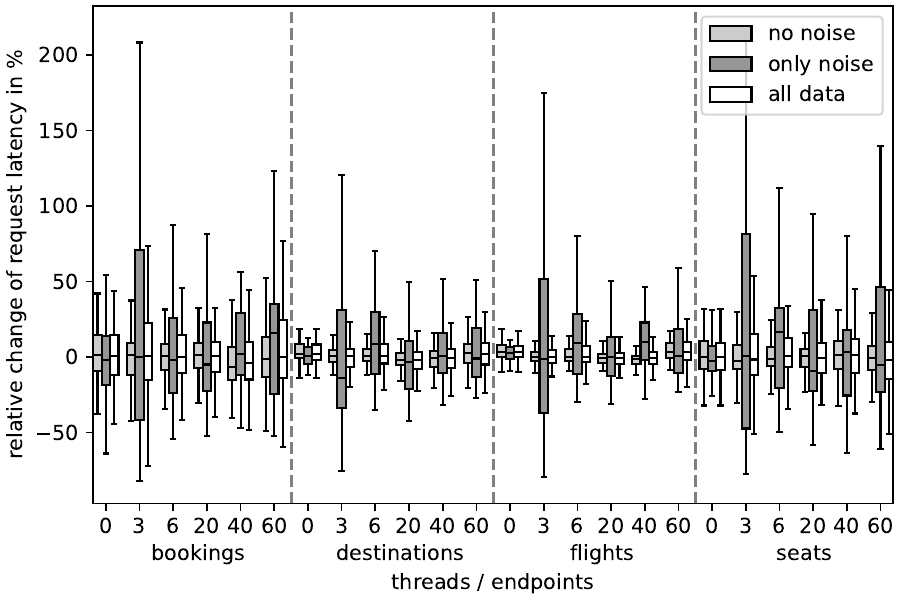} 
    \caption{Boxplots for all results of \emph{Experiment 3 -- Docker containers}.
    For each endpoint, we show the distribution of relative changes in request latency throughout the experiment for all configurations.
    We also show the distribution outside the noise generation (\emph{no noise}), during noise generation (\emph{only noise}), and all result data together (\emph{all data}).}
    \label{fig:experiment-docker}
\end{figure}

\begin{figure}[ht]
    \centering
    \includegraphics[width=\columnwidth]{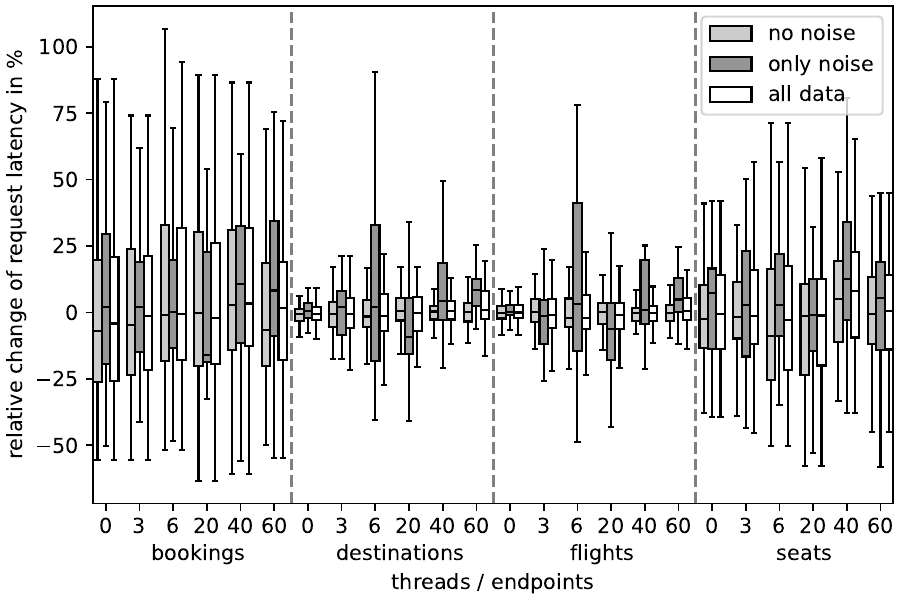} 
    \caption{Boxplots for all results of \emph{Experiment 4 -- Firecracker MicroVMs}.
    For each endpoint, we show the distribution of relative changes in request latency throughout the experiment for all configurations.
    We also show the distribution outside the noise generation (\emph{no noise}), during noise generation (\emph{only noise}), and all result data together (\emph{all data}).}
    \label{fig:experiment-microvm}
\end{figure}

For each experiment, we provide results in the form of boxplots of the resulting distributions of relative changes in request latency that were measured during the experiments (see \cref{fig:experiment-baseline,fig:experiment-core-isolation,fig:experiment-docker,fig:experiment-microvm}).
We show the distributions for each of the four endpoints of the \texttt{flight-booking-service} for all six configurations of CPU noise threads.
Each distribution is further split into the following categories: (1) \enquote{no noise}, which only includes data outside of the phase with the noise generator active, (2) \enquote{only noise}, which only includes data inside the phase with the noise generator active, and (3) \enquote{all data}, which includes all data from the particular experiment.
For \emph{0 threads}, there is no noise generation even during the noise phase.
A potential benchmarking practitioner would only observe \enquote{all data}, which might include strong impact of noise.
Since all experiments were A/A tests, the relative change should always be 0\%.
The distributions of relative changes can therefore give us insight into whether that observation is disturbed by the noise generation.

\emph{Experiment 1 -- Baseline.}
\cref{fig:experiment-baseline} shows the results of the baseline experiment.
We observe that for all four endpoints, the boxplots mostly correctly show a median close to 0\%, but the distributions for \enquote{only noise} are much wider (especially notable for \emph{destinations} and \emph{flights}).
Further comparing the medians of all three boxplots per experiment, the \enquote{no noise} and \enquote{all data} medians tend to lie closer together, while the median of \enquote{only noise} is sometimes far off from 0\% (e.g., \emph{seats} at 20 threads).
This indicates that the noise generation strongly degrades the capability to correctly assess performance changes, but that these effects even out with the data that is unaffected by noise.

\emph{Experiment 2 -- cgroups and CPU pinning.}
\cref{fig:experiment-core-isolation} shows the results of the \emph{cgroups} and \emph{CPU pinning} experiment.
In contrast to the baseline experiment, we can see that the distributions during the noise phase are generally closer to the distributions outside the noise phase.
There is, however, a strong outlier at 3 threads of noise for the \emph{destinations}, \emph{flights}, and \emph{seats} endpoints, where the noise effect is much stronger than during all other configurations, which have higher noise generation.
This could be caused by CPU scheduling, which might behave differently for a higher number of threads within the noise process.

\emph{Experiment 3 -- Docker containers.}
\cref{fig:experiment-docker} shows the results of the Docker container experiment.
Here, we can observe the same effect for 3 threads of noise as in experiment 2, however, the distributions during the noise phase remain wider in all configurations.
This suggests that, unlike for regular \emph{cgroups} and \emph{CPU pinning}, the resource allocation for Docker containers remains more contested, despite Docker using the same mechanisms for resource assignment.

\emph{Experiment 4 -- Firecracker MicroVMs.}
\cref{fig:experiment-microvm} shows the results of the Firecracker MicroVM experiment.
Here, we can observe the same effect at 6 threads of noise that we encountered in experiment 2 and 3 at 3 threads of noise.
Overall, Firecracker MicroVMs show lower effects due to noise (note the different y-axis scaling).
This suggests that the increased process isolation mitigates more of the effects due to noise.

\begin{table*}[tbp]
    \caption{Overview of all measured false positives (FP) across all experiments.
    During each experiment, we compared\break $\bm{6 \; \text{configurations} \times 4 \; \text{endpoints} = 24}$ observations.
    We show the number of observations, split by \enquote{no noise}, \enquote{only noise}, and \enquote{all data}, which were reported as having a \emph{performance change} (i.e., v2 is either slower or faster than v1) by the CI overlap and Wilcoxon rank-sum test.
    All $\bm{24 - n}$ unlisted observations were correctly reported as \emph{no performance change}.}
    \label{tab:false positives}
    \centering
    \begin{tabular}{lcccccc}
        \toprule
        \multirow{2}{*}[-2pt]{Experiment} & \multicolumn{3}{c}{FP by CI Overlap} & \multicolumn{3}{c}{FP by Wilcoxon Rank-Sum Test} \\
        \cmidrule(l{2pt}r{2pt}){2-4}\cmidrule(l{2pt}r{2pt}){5-7}
        & no noise & only noise & all data & no noise & only noise & all data \\
		\midrule
        1 -- Baseline & 0 & 3 & 1 & 1 & 4 & 1 \\
        2 -- cgroups + CPU Pinning & 0 & 0 & 0 & 1 & 0 & 1 \\
        3 -- Docker Containers & 4 & 1 & 3 & 6 & 3 & 3 \\
        4 -- Firecracker MicroVMs & 0 & 2 & 0 & 0 & 4 & 0 \\
        \bottomrule
    \end{tabular}
\end{table*}

\emph{False Positives in Performance Change Detection.}
Now, we evaluate how state-of-the-art performance change detection techniques are influenced by the noise generation.
We calculate significant performance changes according to CI overlap and the Wilcoxon rank-sum test, as explained in \cref{sec:perf-change-detection}, and show all false positives (FP) measured in all experiments in \cref{tab:false positives}.
More detailed result tables for all experiments are available in the online appendix of our replication package.

When comparing experiments 2 and 4 to the baseline experiment, we can see that the employed isolation techniques strictly decrease FPs.
This is different, however, for experiment 3, i.e., Docker container isolation, where even outside of the noise phase both performance change detection methods show several statistically significant performance changes.
This is in contrast to experiment 2, since Docker also builds on \emph{cgroups} and \emph{CPU pinning} for isolating processes.
This leaves experiment 3 as the only experiment, where performance changes for \enquote{all data} increased compared to the baseline.
As such, we do not recommend to use Docker for isolating processes in benchmarks, and to instead use either of the other isolation strategies.

\section{Discussion}
\label{sec:discussion}

In this section, we discuss the limitations of this study.

\textbf{Narrow Scope of Simulated Noise:}
The experiments in this study focused solely on CPU-based contention using thread saturation.
While this form of interference is highly relevant in multi-tenant cloud environments, it does not represent the full spectrum of real-world resource contention.
Cloud-native applications often face performance degradation due to memory pressure, I/O bottlenecks, or network latency.

\textbf{Generalizability Across Cloud Platforms:}
The evaluation was conducted exclusively on Google Cloud using\break \texttt{n2-custom-6-4096} instances.
However, isolation strategies may behave differently across cloud providers due to variations in hypervisor technologies (e.g., KVM on Google Cloud versus Nitro on AWS), CPU architectures (Intel vs. AMD), and resource allocation policies.
For example, Firecracker MicroVMs are optimized for AWS Nitro hardware, and their performance characteristics might vary when deployed on KVM-based hosts like Google Cloud.
The fixed VM type also limits generalizability, as it remains unclear how isolation strategies scale with different instance sizes or resource configurations.

\textbf{Application-Specific Limitations:}
All experiments were conducted using a single application: the \texttt{flight-booking-service} testbed.
While its architecture represents common microservice patterns, and it includes both stateful (write) and stateless (read) operations, it is nonetheless a simplified service with minimal external dependencies and in-memory storage.
This limits its resemblance to complex production systems, which often involve distributed data stores, API integrations, and asynchronous processing.
Isolation strategies may behave differently in applications with more intensive disk I/O, network activity, or concurrency challenges.
Additionally, the nature of duet benchmarking itself may not be suitable for all system types.
Applications with multiple necessary components (e.g., distributed applications) may not work well with isolated execution.
Duet benchmarking assumes independent execution of multiple SUT instances, which is not always feasible in real-world architectures.
The reliance on Go as the implementation language may also introduce language-specific artifacts due to Go's concurrency model and garbage collection behavior.
Broader validation across different languages and system architectures is needed to verify the generalizability of these findings.

\textbf{Limitations of Application Workload:}
The benchmark workload involved 50 VU executing scripted interactions.
While this setup generated measurable performance differences, it does not capture the scale or variability of real-world traffic.
The absence of traffic spikes, irregular user patterns, or bursty requests likely underrepresents tail latency and peak stress scenarios.
As a result, the impact of isolation on outlier behavior (e.g., high-percentile latency) may be underestimated.

\textbf{Statistical and Methodological Constraints:}
The performance comparisons relied on median response ratios and bootstrap confidence intervals.
While statistically sound, this method involved aggregating data to one median value per second, which may obscure high-frequency fluctuations or fine-grained instability.
Additionally, the 99\% confidence level, while conservative, may have overlooked subtle but relevant changes.

\section{Related Work}
\label{sec:rel_work}

\paragraph{Detecting and Quantifying Performance Changes}
Benchmarking requires repeatability and consistent results under identical conditions.
However, this is difficult in cloud environments due to dynamic resource allocation, hardware heterogeneity, and interference from noisy neighbors~\cite{Bulej_2020,book_bermbach2017_cloud_service_benchmarking,schad2010runtime,uta2020big}.
A common approach is to use RMIT to repeat and shuffle iterations of a particular workload~\cite{AbediOld,abedi2017conducting}.
This increases the repeatability of results in the presence of noise, but also increases the experiment time and, therefore, cost.
On the platform side, cloud providers can use host-level strategies like interference-aware scheduling and VM migration.
For instance, Delimitrou and Kozyrakis propose an interference-aware scheduler that optimizes server consolidation~\cite{delimitrou2013paragon}, while Roytman et al.\ use cache access behavior to guide VM placement~\cite{roytman2013algorithm}.
When using public clouds, however, practitioners need to use special analysis methods to obtain good results on performance regressions.
Amannejad et al.\ propose \emph{Collaborative Response Time Estimation}, a machine learning approach using collaborative filtering to predict expected transaction times~\cite{virtissue}.
Deviations from these estimates signal performance interference and enable dynamic mitigation strategies like load redistribution.
Daly et al.\ apply the E-Divisive means algorithm to identify changepoints in continuous integration systems to enhance performance regression detection~\cite{daly2020change}.
Fleming et al.\ build on this by replacing the permutation-based testing in E-Divisive with a Student's $t$-test in their tool \emph{Hunter}, which improves accuracy and runtime \cite{fleming2023hunter}.
Ordozgoiti et al.\ employ deep CNNs to identify noisy neighbors from time-series performance metrics, achieving high classification accuracy in real cloud environments~\cite{deep_detecting}.
There also exist approaches to determine the optimal execution time for a benchmarking experiment in the presence of noise~\cite{japke_uoptime_2025,he_statistics_2019,laaber_dynamically_2020,Grambow2021,japke2025towards}

\paragraph{Performance Comparisons of Virtualization and Isolation Methods}
Ghatrehsamani et al.\ compare CPU pinning effects across containers, VMs, and bare-metal platforms~\cite{ghatrehsamani2020art}.
Their results show that CPU pinning reduces overhead significantly, especially in I/O-intensive applications.
Containers with more pinned cores exhibited better performance and lower overhead, and pinning helped mitigate resource tracking overhead within containerized environments.
Podzimek et al.\ study CPU pinning in colocated workloads and find that unconventional pinning configurations improve energy efficiency, particularly under partial background loads~\cite{podzimek2015analyzing}.
Their findings support dynamically adapting pinning strategies based on workload characteristics and system load.
Docker's performance has also been widely studied, e.g., comparing Docker against bare-metal systems~\cite{saha2018evaluation,casalicchio2017measuring}.
Grambow et al.\ focus on Docker's impact on database benchmarking~\cite{grambow2019safe}.
Their findings suggest Docker can introduce unpredictable latency and throughput degradation, particularly for write-heavy workloads.
In the context of MicroVMs, Wang compares Firecracker with Docker and QEMU-based virtual machines across multiple metrics~\cite{wang2021can}.
Firecracker offered a middle ground: better isolation and security than containers but without the full overhead of traditional VMs.
However, Firecracker did not consistently outperform Docker, especially in startup time and I/O throughput.
Similarly, Hebbar et al.\ use Sysbench and RAMSpeed to benchmark Firecracker, Docker, and bare-metal systems~\cite{hebbar2022performance}.
Their results indicate that Firecracker matches Docker in CPU and memory performance, but lags in file I/O.
They conclude that while Firecracker fulfills its design goals in terms of isolation, further improvements are needed to bridge remaining performance gaps.

\section{Conclusion}
\label{sec:conclusion}

In this work, we developed a noise generating application, and ran benchmarking experiments comparing three different process isolation techniques.
We found that noise strongly affects performance measurements, but when analyzing the relative change of both SUT versions, duet benchmarking mostly provides correct results.
Furthermore, we found that process isolation still decreased the amount of false positives, except when using Docker containers.

\balance

\bibliographystyle{ACM-Reference-Format}
\bibliography{refs.bib}

\end{document}